\documentclass{appolb}

\usepackage{graphicx}
\usepackage{color}
\usepackage{slashed}
\usepackage[normalem]{ulem}
\usepackage{amsmath,amssymb}

\usepackage{bbm}                                                  

\newcommand{\nc}{\newcommand}
\nc{\beq}{\begin{equation}}  \nc{\eeq}{\end{equation}}
\nc{\bea}{\begin{eqnarray}}  \nc{\eea}{\end{eqnarray}}
\nc{\baa}{\begin{array}}     \nc{\eaa}{\end{array}}
\nc{\bit}{\begin{itemize}}   \nc{\eit}{\end{itemize}}
\nc{\ben}{\begin{enumerate}} \nc{\een}{\end{enumerate}}
\nc{\bce}{\begin{center}}    \nc{\ece}{\end{center}}
\nc{\bpm}{\begin{pmatrix}}   \nc{\epm}{\end{pmatrix}}
\nc{\bvt}{\begin{verbatim}}  \nc{\evt}{\end{verbatim}}
\nc{\non}{\nonumber} 

\def\half{\frac12}	

\def\to{\rightarrow}

\def\gesim{\gtrsim}

\def\boldoverdot{\,{\raise6pt\hbox{\bf.}\!\!\!\!\>}}

\def\gcal{{\cal G}}

\def\lcal{{\cal L}}

\def\zBB{{\mathbbm Z}}
\usepackage{bm}
\def\diag{\hbox{\diag}}

\def\gev{\;\hbox{GeV}}
\def\tev{\;\hbox{TeV}}

\def\doubleundertext#1{
{\undertext{\vphantom{y}#1}}\par\nobreak\vskip-\the\baselineskip\vskip4pt%
\undertext{\hbox to 2in{}}}
\def\inbox#1{\vbox{\hrule\hbox{\vrule\kern5pt
     \vbox{\kern5pt#1\kern5pt}\kern5pt\vrule}\hrule}}
\def\sqr#1#2{{\vcenter{\hrule height.#2pt
      \hbox{\vrule width.#2pt height#1pt \kern#1pt
         \vrule width.#2pt}
      \hrule height.#2pt}}}

\def\today{\ifcase\month\or
  January\or February\or March\or April\or May\or June\or
  July\or August\or September\or October\or November\or December\fi
  \space\number\day, \number\year}
\def\pmb#1{\setbox0=\hbox{#1}%
  \kern-.025em\copy0\kern-\wd0
  \kern.05em\copy0\kern-\wd0
  \kern-.025em\raise.0433em\box0 }

\def\pmbb#1{\setbox0=\hbox{#1}%
  \kern-.02em\copy0\kern-\wd0
  \kern.04em\copy0\kern-\wd0
  \kern-.02em\raise.03464em\box0 }
\def\sumprime_#1{\setbox0=\hbox{$\scriptstyle{#1}$}
  \setbox2=\hbox{$\displaystyle{\sum}$}
  \setbox4=\hbox{${}'\mathsurround=0pt$}
  \dimen0=.5\wd0 \advance\dimen0 by-.5\wd2
  \ifdim\dimen0>0pt
  \ifdim\dimen0>\wd4 \kern\wd4 \else\kern\dimen0\fi\fi
\mathop{{\sum}'}_{\kern-\wd4 #1}}
\def\vp{{\varphi}}
\def\lx{\lambda_x}

\def\mvp{m_\varphi}

\def\non{\nonumber}

\def\yuk{g_\nu}

\def\gt{\zeta}

\begin{document}
\title{Constraints on Two-Component Dark Matter%
\thanks{Presented at the XXXVII International Conference of Theoretical Physics "Matter to the Deepest 2013"}%
}
\author{
Subhadittya Bhattacharya
\address{ Department of Physics, University of California, Riverside, CA  92521, USA}
Aleksandra Drozd, Bohdan Grzadkowski
\address{\,Faculty of Physics, University of Warsaw, 00-681 Warsaw, Poland}
\\
Jose Wudka
\address{ Department of Physics, University of California, Riverside, CA  92521, USA}
}
\maketitle
\begin{abstract}
We study 'Higgs Portal' 2-component Dark Matter scenario with two interacting cold Dark Matter (DM) candidates: a neutral scalar singlet ($\vp$) and a neutral Majorana fermion ($\nu$). The relic abundance of $\nu$ and $\vp$ is found assuming thermal DM production and solving the Boltzmann equations. We scan over the parameter space of the model to determine regions 
consistent with the WMAP data for DM relic abundance and the XENON100 direct detection limits for the DM-nucleus cross section.
\end{abstract}


\section{Introduction}
There are many reasons to consider physics beyond the Standard Model of Elementary Interactions (SM) - one of them is to provide stable, massive, neutral particles that might play a role of Dark Matter (DM) \cite{Bertone:2004pz}. An enormous amount of work has been done by theoreticians in this direction, considering many types of models, most of which contain a single particle beyond the SM that might be considered  a DM candidate. 

However, there is no reason to limit our search to single-component DM scenarios only. 

 Multi-component DM hypothesis seems particularly enticing considering how complex is the structure of SM matter accounting for only $\sim 5\%$ of energy density of the Universe.
 There have already been studies of multi-component DM in the literature (see \cite{Profumo:2009tb, Drozd:2011aa, Daikoku:2011mq, Huh:2008vj})
dealing with - among others - discrepancies between different results of DM searches or solving the 'core/cusp problem'  \cite{Medvedev:2013vsa}.

Here we would like to investigate a scenario where DM consists of two species --
a singlet scalar ($\vp$) and a singlet neutral {\em Majorana} fermion ($\nu$)
(that we will refer to as a ''neutrino''). The scalar DM field in this model interacts with the 
SM through the Higgs field, while the fermionic DM does not couple directly to the SM.  In our study we will concentrate on finding agreement with the WMAP data and XENON 100 results \cite{Aprile:2012nq}.

\section{The model}
Our model contains three new particles, all SM singlets: a real scalar $ \vp $, and two {\em majorana} fermions $\nu_h$ and $\nu$, with only {\em one} of  the fermions contributing to the DM relic density. In order to ensure stability of DM candidates we will assume that the dark sector is invariant under some global symmetry group  $ \gcal $ under which all the extra fields transform non-trivially, while all SM particles are $ \gcal $-singlets. 
For simplicity we choose $ \gcal=\zBB_2 \times \zBB_2 $. The dark sector transforms  under  $ \gcal $ as follows:
\beq
\nu_h \sim [-,+] \quad
\nu   \sim [+,-] \quad
\vp    \sim [-,-]
\eeq
The most general, gauge- and $\gcal$-symmetric and renormalizable Lagrangian is:
\beq
\begin{split}
\lcal_{\rm scal} &= \frac{1}{2} \partial_{\mu} \varphi \partial^{\mu} \varphi + 
D_{\mu}H^{\dagger} D^{\mu} H 
- V(H, \varphi) \, ,
 \\
V(H, \varphi) &= 
- \mu_{H}^2 H^{\dagger} H 
+ \lambda_{H} (H^{\dagger} H)^2 
\\
& \quad
+ \frac{1}{2}  \mu_{\varphi}^2 \varphi^2
+ \frac{1}{4!} \lambda_{\varphi} \left(\varphi^2 \right)^2
 +  \lambda_{x} H^{\dagger} H \varphi^2 \,.
\end{split}
\label{pot}
\eeq
where $H$ is the SM $SU(2)$  Higgs isodoublet.

We require that the potential breaks spontaneously the electroweak symmetry via 
non-zero vacuum expectation value of the Higgs doublet $\left<H\right> = (0, v/\sqrt{2})$, 
$v = 246 \gev$. Since we also require the $\gcal$ symmetry to remain unbroken, we assume that $\mu_\varphi^2 > 0$, which is a sufficient condition.  Note that $\left<\vp\right> = 0$ implies that there is no mass-mixing between $\vp$ and $H$, so that the existing collider limits on the Higgs properties are not modified.
After the symmetry breaking, the physical scalars acquire masses 
${m_H}^2 = - \mu_{H}^2 + 3 \lambda_{H} v^2 =2 \mu_{H}^2$ 
and $m_{\varphi}^2 = \mu_{\varphi}^2 + \lambda_{x} v^2 $. 

The part of the DM Lagrangian involving fermions reads:
\beq
\lcal = 
\half \overline{\nu_h} \, i \!\! \not\!\partial \, \nu_h + 
\half \overline{\nu} \, i \!\! \not\!\partial \, \nu -
\frac12 \nu_h^T C \nu_h M_h - \frac12 \nu^T C  \nu m_\nu +
\yuk \vp \, \overline{\nu_h} \nu.
\label{lag2}
\eeq


 In order to ensure stability of $\nu$ and $\vp$ 
we assme that $ M_h > m_\nu + m_\vp $, which allows for the decay of $ \nu_h \to \vp \nu $. 
The tree level reactions relevant for the evolution of DM are
$ \vp\vp \leftrightarrow \rm SM,SM$ and $ \vp\vp \leftrightarrow \nu\nu $. 

\subsection{Restrictions on the parameter space}
In the following we will fix $M_h$ at the smallest value that ensures 
the fast decay of $\nu_h$, so
we will effectively deal with only four parameters: $\mvp, m_\nu, \lx$ and $g_\nu$.
Our goal is to constrain the parameters taking into account all available theoretical restrictions: vacuum stability, unitarity, perturbativity, triviality of the scalar sector (for discussion see \cite{ Drozd:2011aa}). In this paper we will concentrate on constraining our model according to the WMAP bounds on DM relic abundance and the results of XENON 100 experiments.

\section{Dark Matter Density and the Boltzmann Equation}
\label{BE}

 To test the model against the relic density constraint derived from WMAP we start with formulating and solving the two Boltzmann equations (BEQ) that govern the cosmological 
evolution of  the DM neutrinos ($\nu$) and scalar singlets ($\vp$). 
We assume kinetic equilibrium and neglect possible effects of quantum statistics. The BEQs (including tree-level interactions only) read: 
\beq
\begin{split}
 \dot{n}_{\vp}+3H n_{\vp} & =
 -\langle \sigma_{ \vp \vp \to SM \, SM} v \rangle \left(n_{\vp}^2- n_{\vp}^{EQ} {}^2\right) +
\\ & \quad
- \left(
\langle \sigma_{ \vp \vp \to \nu \nu} v \rangle 
n_{\vp}^2- 
\langle \sigma_{\nu \nu \to \vp \vp} v \rangle n_{\nu}^2
\right) \\
 \dot{n}_{\nu}+3H n_{\nu} & = 
- \left(
\langle \sigma_{\nu \nu \to \vp \vp} v \rangle n_{\nu}^2 -
\langle \sigma_{ \vp \vp \to \nu \nu} v \rangle 
n_{\vp}^2
\right)
\end{split}
\label{beq2}
\eeq
with $n_X$ - the number density of $X=\nu, \vp$, a dot - time derivative,  $H$ - the Hubble parameter. $\langle \sigma_{ X X \to Y Y} v \rangle$ is a thermally averaged cross section (see \cite{Gondolo:1990dk})
The equilibrium density $n_X^{EQ}$ is related to the equilibrium phase space density $\tilde{f}_X^{EQ}$ as follows:
\beq
n_X^{EQ} =  \int \frac{\gt_X d^3 p}{(2 \pi)^3 2E}  \tilde{f}_X^{EQ},    \hspace{.1 cm}   \tilde{f}_X^{EQ}  = \frac{1}{e^{E/T}\pm1}, 
\eeq
The chemical potential for the above vanishes, and $\pm$ refers to fermions and bosons, respectively. It is important to remember that the thermally averaged cross sections that appear in (\ref{beq2}) are not independent,
\beq
\langle \sigma_{\nu \nu \to \vp \vp} v \rangle = \left(\frac{n_{\vp}^{EQ}}{n_{\nu}^{EQ}}\right)^2 \langle \sigma_{ \vp \vp \to \nu \nu} v \rangle \,.
\label{csrelation}
\eeq

\subsection{Numerical solutions to the BEQs}

\begin{figure}[htb]
\centerline{
\includegraphics[height = 4.2 cm]{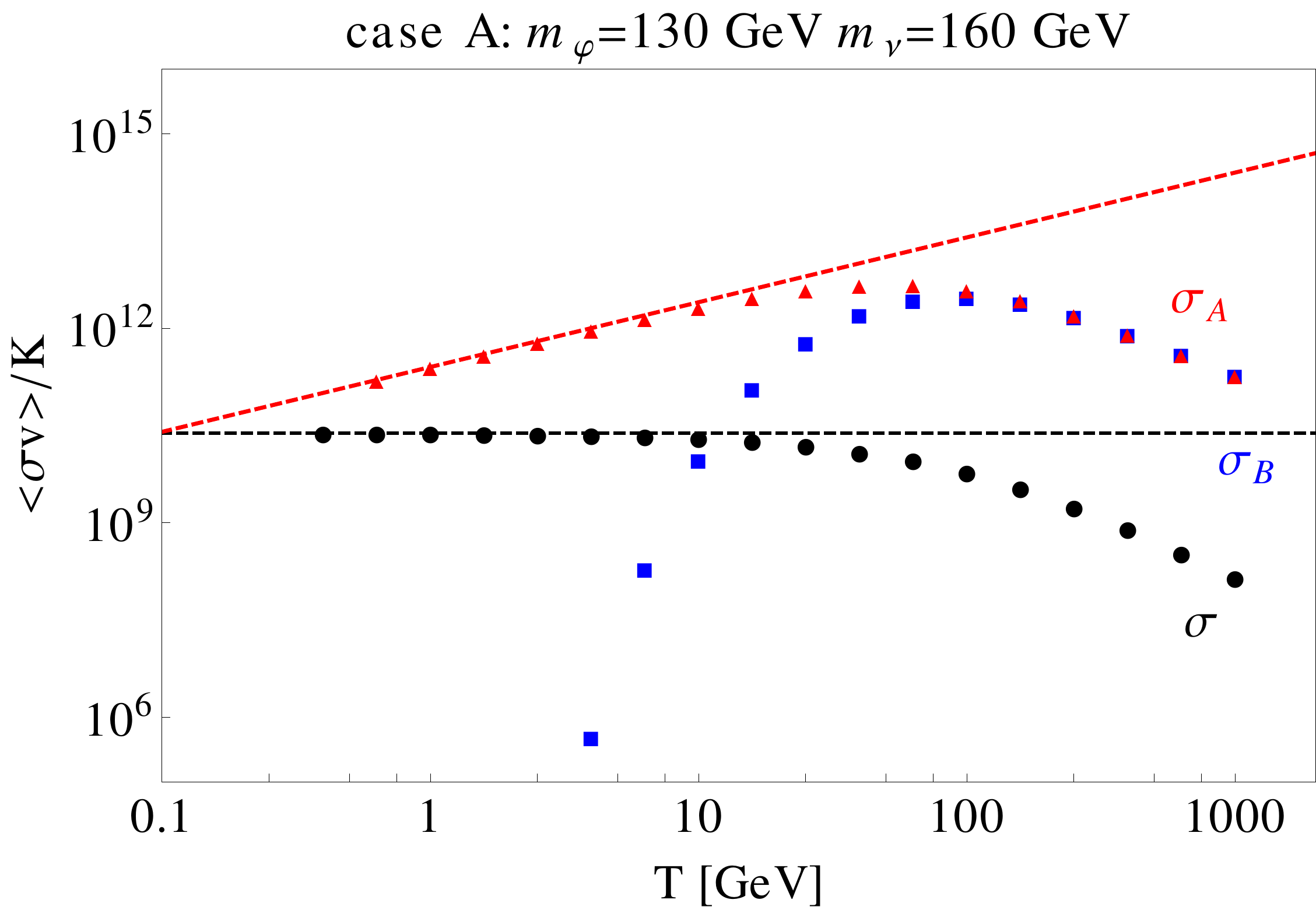}
\includegraphics[height = 4.2 cm]{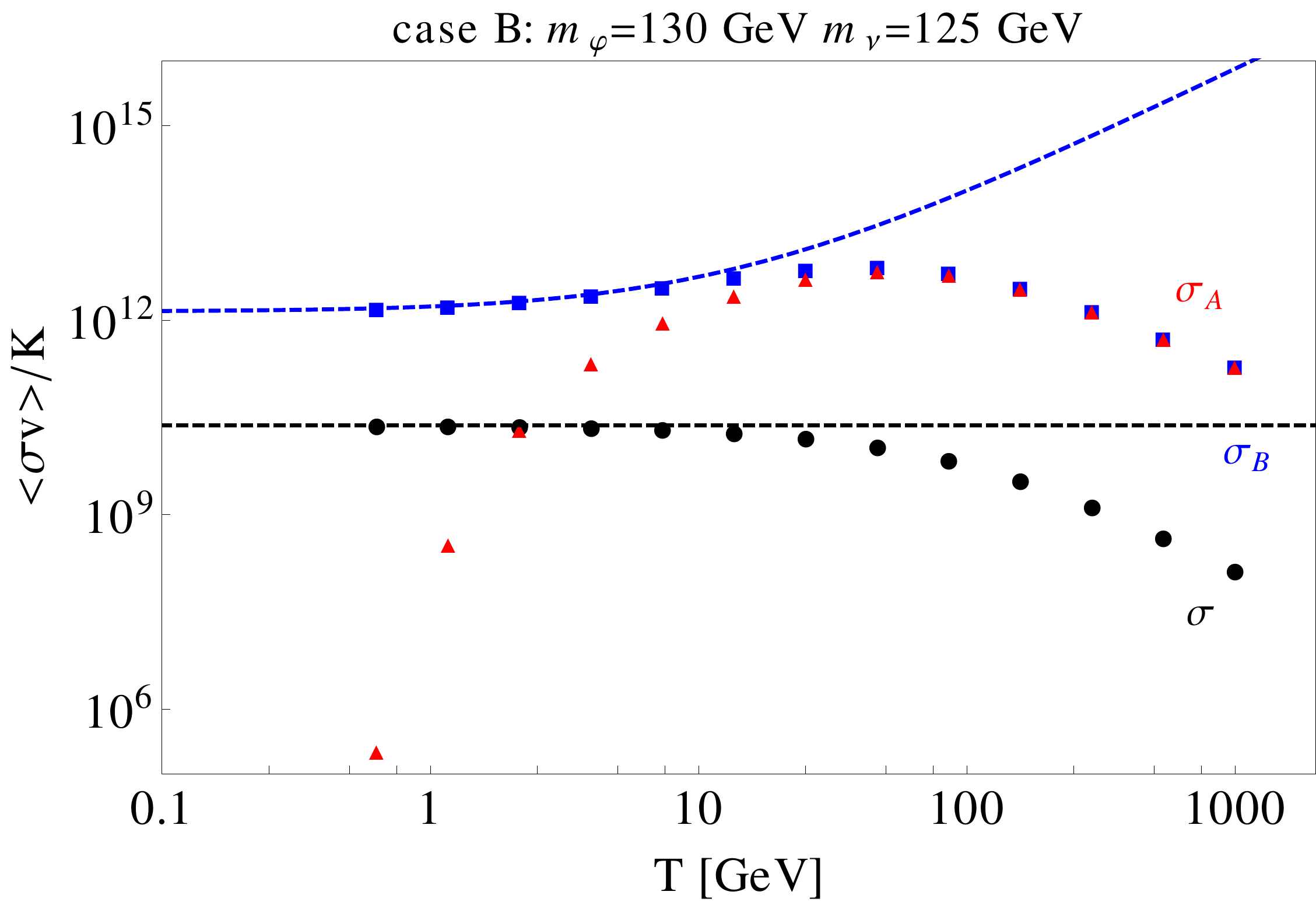}}
\caption{Thermally averaged cross sections 
$\sigma \equiv $ $\langle \sigma_{ \vp \vp \to SM \, SM} v \rangle/K$ (black points);
$\sigma_B \equiv $$ \langle \sigma_{ \vp \vp \to \nu \nu} v \rangle/K$ (green points);
$\sigma_A \equiv $$\langle \sigma_{ \nu \nu \to \vp \vp} v \rangle/K$ (red points),
as functions of $T$ (in GeV),
for $\lx = .1$ and $g_\nu = 2.5$, masses are specified in the plots.}
\label{fig_cs_AB}
\end{figure}

For the purposes of solving equations (\ref{beq2}) numerically we approximate the thermally averaged cross sections in (\ref{beq2}) by first order terms in the expantion around $T~=~0$, see fig.~\ref{fig_cs_AB}. For each point in the parameter space of the model we approximate the low temperature dominant cross section (whether it is $\langle \sigma_{\nu \nu \to \vp \vp} \rangle$ or $\langle \sigma_{ \vp \vp \to \nu \nu} v \rangle$, depends on the mass hierarchy between $\vp$ and $\nu$) by its leading terms in the temperature expansion, and use relation (\ref{csrelation}) for the other one. As a result equations (\ref{beq2}) are simplified as follows:
\bea
f_{\vp}^\prime &=&
\sigma\left[f_{\vp}^2- {f_{\vp}^{EQ} }^2\right] + 
\sigma_A \left[\left(\frac{f_{\nu}^{EQ}}{f_{\vp}^{EQ}}\right)^2 f_{\vp}^2 -  f_{\nu}^2 \right] 
\label{BEAvp} \\
f_{\nu}^\prime &=& 
\sigma_A \left[f_{\nu}^2 - \left(\frac{f_{\nu}^{EQ}}{f_{\vp}^{EQ}}\right)^2 f_{\vp}^2\right],\,
\sigma_A \propto T  
\label{BEAnu}
\eea
for  $m_\nu>\mvp$ and
\bea
f_\vp^\prime &=&
\sigma  \left[f_{\vp}^2- f_{\vp}^{EQ} {}^2\right] 
+\sigma_B \left[
 f_{\vp}^2- 
\left(\frac{f_{\vp}^{EQ}}{f_{\nu}^{EQ}}\right)^2 f_{\nu}^2
\right] \label{BEB1}\\
f_\nu^\prime &=& 
\sigma_B \left[
\left(\frac{f_{\vp}^{EQ}}{f_{\nu}^{EQ}}\right)^2f_{\nu}^2 - f_{\vp}^2
\right],\,
\sigma_B = \sigma_B^0 + \sigma_B^1 T + \sigma_B^2 T^2
\label{BEB2}
\eea
for  $m_\nu<\mvp$. The relative errors of this approximation are: 
$\delta^A_\vp \simeq 2.3\% $, $\delta^A_\nu \simeq 1.4\% $, $\delta^A_\vp \simeq 6.3\% $, $\delta^A_\nu \simeq 2.6\% $, where A and B stands for $m_\nu>\mvp$ and $m_\nu<\mvp$, respectively.

We perform a scan over the parameter space in the following ranges: 
\beq
\mvp, m_\nu \in (10 \mathrm{\,GeV}, 10  \mathrm{\,TeV}), \,\, \lambda_x \in (10^{-3}, 4\pi), \,\, g_\nu  \in (0.1, 4\pi) \non
\eeq
solving the BEQs and fitting the relic abundance of $\vp$ and $\nu$ to the WMAP results \cite{Hinshaw:2012aka}. We assume that abundance for both scalar and neutrino DM components must be below WMAP $3\sigma$ upper limit,   
\beq
\Omega_{\vp}+\Omega_{\nu}\leq\Omega_{WMAP} = 0.1138+3*0.0045\,,
\eeq
assuming that in general there might be other components of DM, non-interacting with SM other than gravitationally, that contribute to the DM density, but do not thermalize in the Early Universe. Such components might be responsible for any missing density (if $\Omega_{\vp}+\Omega_{\nu} < 0.1138-3~*~0.0045$).

\begin{figure}[htb]
\centerline{
\includegraphics[height = 3.9 cm]{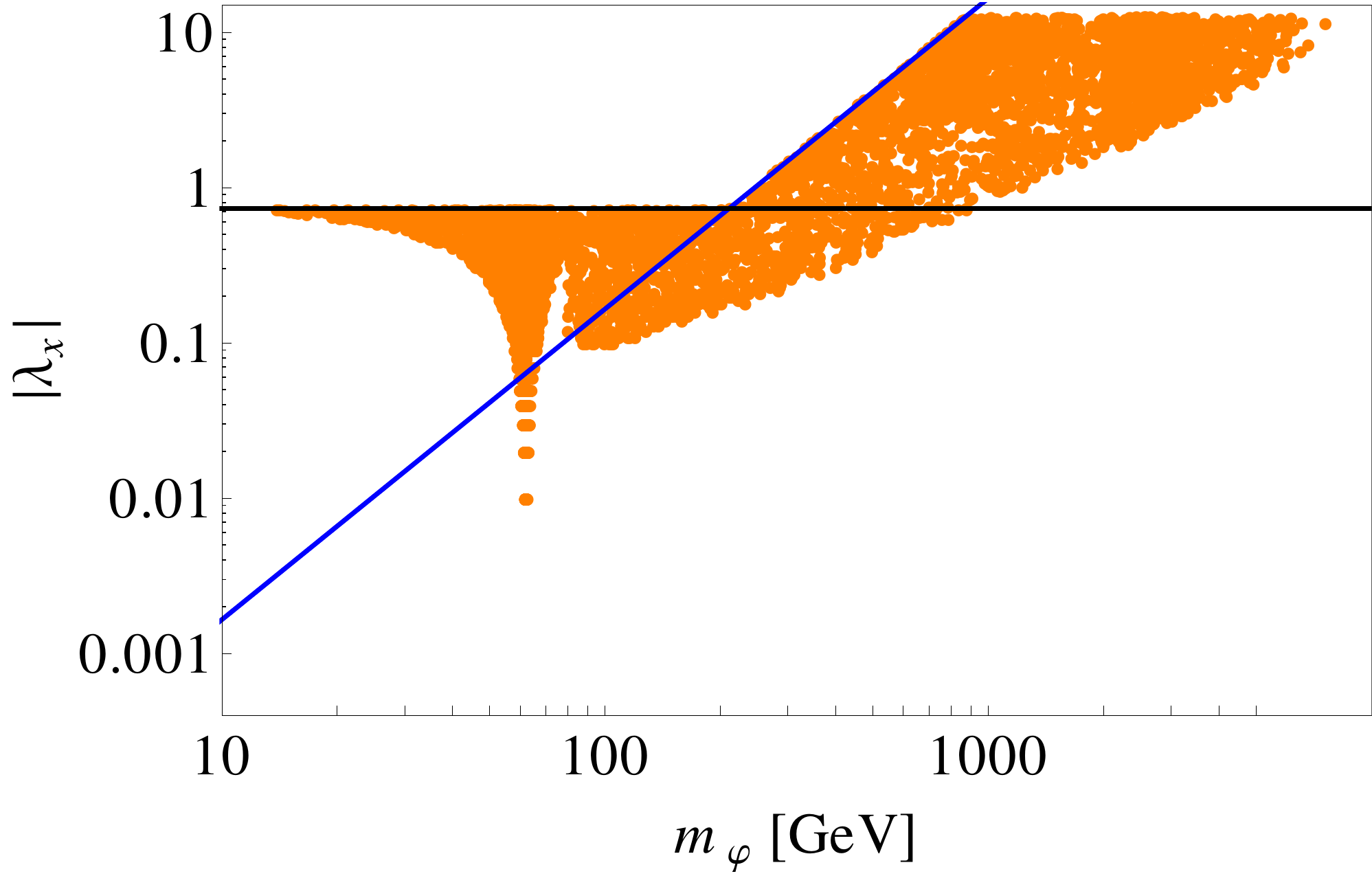}
\includegraphics[height = 4.2 cm]{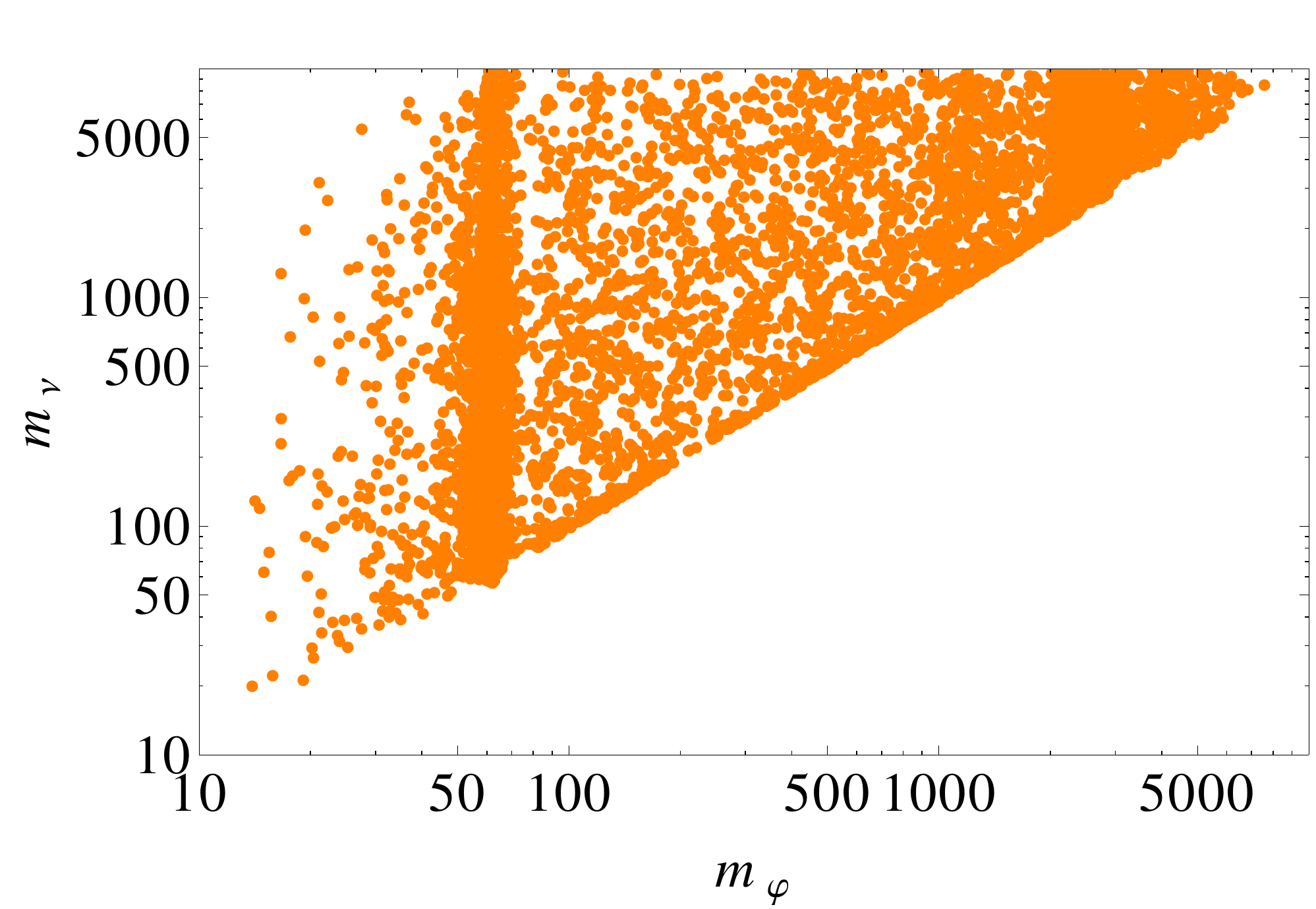}}
\caption{Points that satisfy the upper WMAP bound for the relic abundance, projected into
$(\mvp,|\lx|)$ and $(\mvp,m_\nu)$  plane (left and right plots, respectively).  Blue line is a theoretical limit assuring there is no VEV for the singlet scalars and the scalar mass square is positive, see \cite{Drozd:2011aa}. It is an upper cut on the positive branch of $\lx$, while the black horizontal line is the stability limit that bounds $\lambda_x$ from below, for negative values.}
\label{DD_res} 
\end{figure}

\section{Direct Detection}

%
%

In our model, at the tree level, scattering of DM off nuclei originates from the interaction with the scalar DM component. 
However, DM is often dominated by dark neutrinos in which case loop induced $\nu$ nucleon scattering might be relevant. To compare the prediction for the direct detection cross section obtained for our 2-component DM scenario
with experimental results, one has to remember that the standard limits on DM direct detection usually assume all DM particles to be interacting with SM at the same rate. In our case, this is not true and we will rescale the cross sections by a factor that accounts for the fact that two DM components are present:
\beq
\sigma_{\rm DM-N}^\vp = \frac{n_{\vp}}{n_{\vp}+n_{\nu}} \sigma_{\vp  N}, \,\,
\sigma_{\rm DM-N}^\nu = \frac{n_{\nu}}{n_{\vp}+n_{\nu}} \sigma_{\nu  N}.
\label{dir_DM_vp}
\eeq
For $ \sigma_{\vp  N}, \sigma_{\nu  N}$ see \cite{Drozd:2011aa}.  As seen from the left panel of fig.~\ref{DD_res} the majority of points for the scalar scattering lie above (i.e. are excluded by) the XENON100 lower limit. Points that are below XENON100 can be found in the resonance region $\mvp\simeq m_h/2$, in the middle mass region $\mvp \simeq 130-140\gev$ and for heavy scalar solution $\mvp \gesim 3\tev$. However, in the region of $\mvp \simeq 130-140\gev$ loop corrections in $\sigma_{\rm DM-N}^\nu$ are large enough to also exclude this mass range.

\begin{figure}[htb]
\centerline{
\includegraphics[height = 4.2 cm]{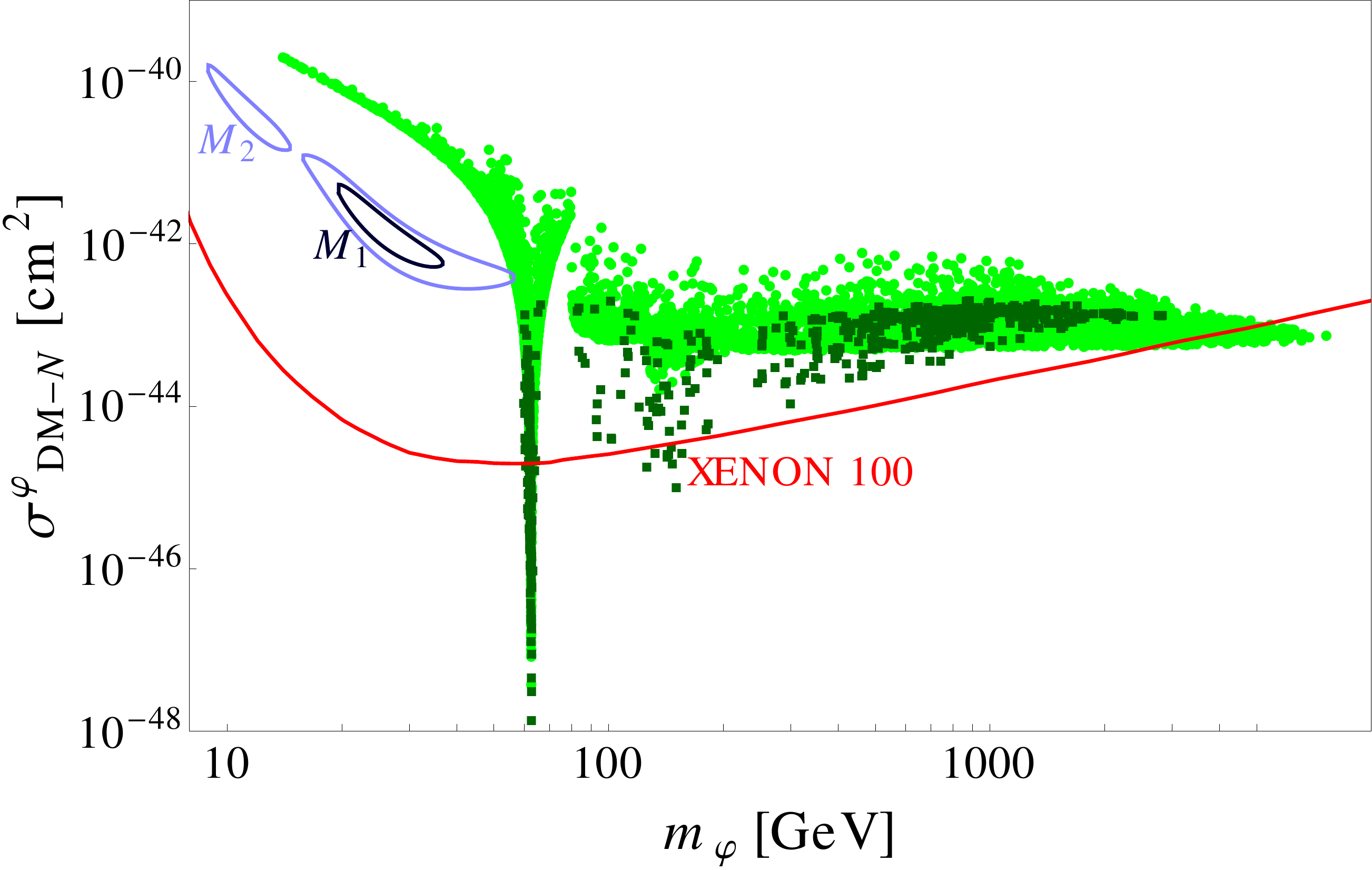}
\includegraphics[height = 4.2 cm]{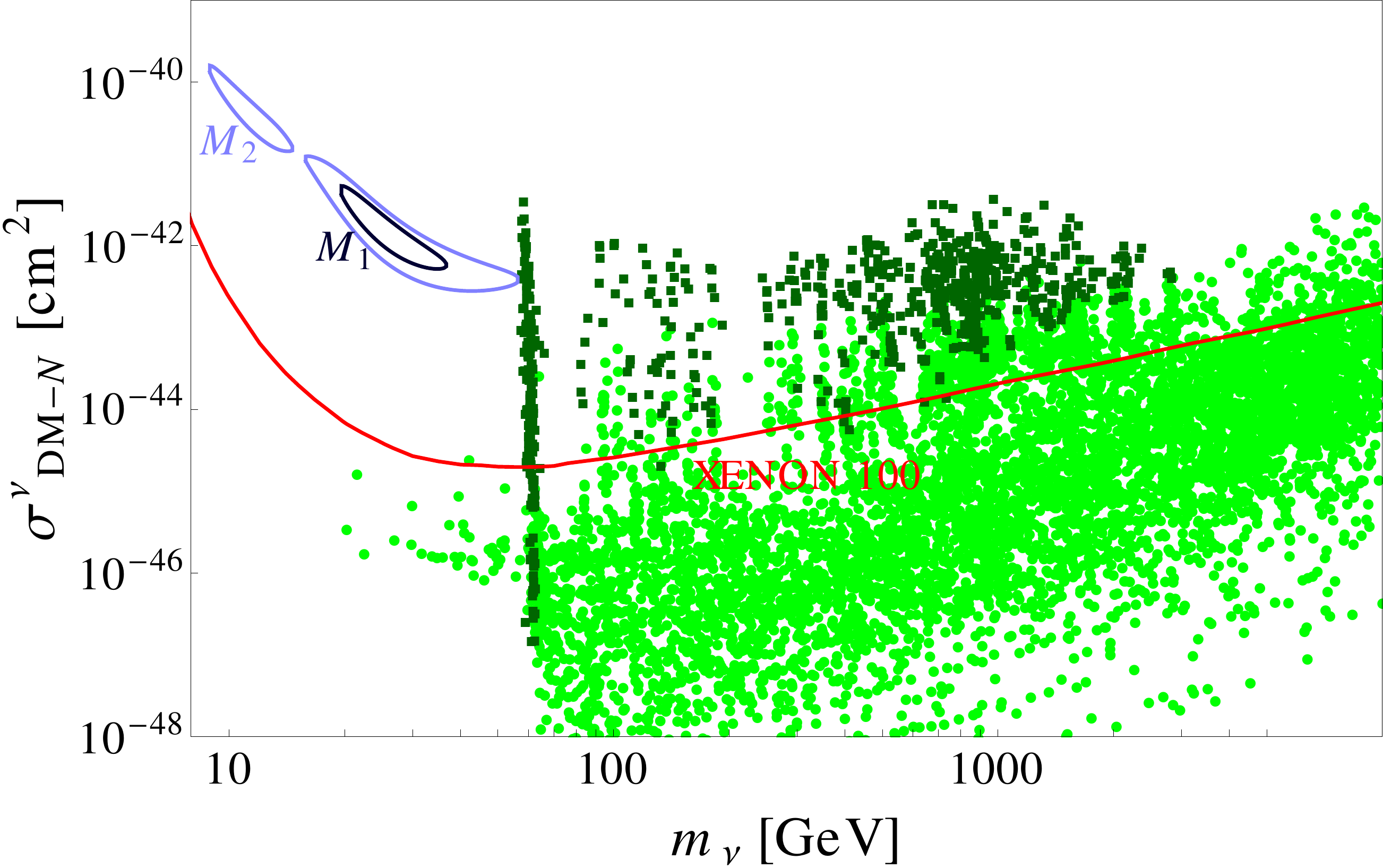}}
\caption{Plot of the cross section $\sigma_{\rm DM-N}^\vp$ as a function of $\mvp$ on the left and $\sigma_{\rm DM-N}^\nu$ as a function of $m_\nu$ on the right. All points satisfy all theoretical constraints and a $3 \sigma$ WMAP upper bound. The other parameters are randomly chosen in the ranges defined in the text (including both signs of $\lx$). Green circles (dark green squares) correspond to case A (case B) solutions. The red line shows the XENON100 data, and the two islands in blue indicate 1 and $2\sigma$ CRESST-II results \cite{Angloher:2011uu}.}
\label{DD_res} 
\end{figure}

\section{Conclusions}
We have discussed the main features of a two-component cold Dark Matter model  composed of a neutral Majorana fermion ($\nu$) and a neutral real singlet ($\vp$). The Boltzmann equations for number densities of $\nu$ and $\vp$ were solved numerically in order to determine regions of parameter space that are consistent with both WMAP and XENON100.

It has been shown that the agreement with the WMAP data requires that neutrinos cannot be substantially lighter than scalars. The XENON100 upper limit in DM-nucleon cross section favours $\mvp \gesim 3 \tev$ with $m_\vp< m_\nu$. We have also found consistent solutions for
$\mvp \simeq m_h/2$.

As a final remark we note that
such a model is difficult to test at the Large Hadron
Collider (LHC). The leading new effect
would be production of scalar DM pairs, with
a signature of missing energy associated with one or more jets.
Such analysis lies beyond the scope of this work.



\begin{thebibliography}{10}

\bibitem{Bertone:2004pz} 
  G.~Bertone, D.~Hooper and J.~Silk,
  Phys.\ Rept.\  {\bf 405}, 279 (2005)
  [hep-ph/0404175].
  
  


\bibitem{Profumo:2009tb}
  S.~Profumo, K.~Sigurdson and L.~Ubaldi,
  JCAP {\bf 0912} (2009) 016
  [arXiv:0907.4374 [hep-ph]].
  G.~B.~Gelmini,
  Nucl.\ Phys.\ Proc.\ Suppl.\  {\bf 138} (2005) 32
  [hep-ph/0310022].
  G.~Duda, G.~Gelmini, P.~Gondolo, J.~Edsjo and J.~Silk,
  Phys.\ Rev.\ D {\bf 67} (2003) 023505
  [hep-ph/0209266].
  G.~Duda, G.~Gelmini and P.~Gondolo,
  Phys.\ Lett.\ B {\bf 529} (2002) 187
  [hep-ph/0102200].
  K.~R.~Dienes and B.~Thomas,
  Phys.\ Rev.\ D {\bf 85}, 083523 (2012)
  [arXiv:1106.4546 [hep-ph]].
  K.~R.~Dienes and B.~Thomas,
  Phys.\ Rev.\ D {\bf 85}, 083524 (2012)
  [arXiv:1107.0721 [hep-ph]].

\bibitem{Drozd:2011aa} 
  A.~Drozd, B.~Grzadkowski and J.~Wudka,
  JHEP {\bf 1204}, 006 (2012)
  [arXiv:1112.2582 [hep-ph]];
  A.~Drozd, B.~Grzadkowski and J.~Wudka,
  Acta Phys.\  Polon.\ B {\bf 42}, 11, 2255 (2011)
  [arXiv:1310.2985 [hep-ph]].
  S.~Bhattacharya, A.~Drozd, B.~Grzadkowski and J.~Wudka,
  arXiv:1309.2986 [hep-ph].

\bibitem{Daikoku:2011mq}
  Y.~Daikoku, H.~Okada and T.~Toma,
  Prog.\ Theor.\ Phys.\  {\bf 126} (2011) 855
  [arXiv:1106.4717 [hep-ph]];
  L.~Bian, R.~Ding and B.~Zhu,
  arXiv:1308.3851 [hep-ph].
  Q.~-H.~Cao, E.~Ma, J.~Wudka and C.~-P.~Yuan,
  arXiv:0711.3881 [hep-ph].
  D.~Feldman, Z.~Liu, P.~Nath and G.~Peim,
  Phys.\ Rev.\ D {\bf 81} (2010) 095017
  [arXiv:1004.0649 [hep-ph]];
  M.~Aoki, M.~Duerr, J.~Kubo and H.~Takano,
  arXiv:1207.3318 [hep-ph];
  J.~Heeck and H.~Zhang,
  arXiv:1211.0538 [hep-ph];
  K.~M.~Zurek,
  Phys.\ Rev.\ D {\bf 79} (2009) 115002
  [arXiv:0811.4429 [hep-ph]].
  Y.~Tomozawa,
  Int.\ J.\ Mod.\ Phys.\ A {\bf 23} (2008) 4811
  [arXiv:0806.1501 [astro-ph]].
  M.~Malekjani, S.~Rahvar and D.~M.~Z.~Jassur,
  New Astron.\  {\bf 14} (2009) 398
  [arXiv:0706.3773 [astro-ph]].
  M.~Heikinheimo, A.~Racioppi, M.~Raidal, C.~Spethmann and K.~Tuominen,
  Nucl.\ Phys.\ B {\bf 876}, 201 (2013)
  [arXiv:1305.4182 [hep-ph]].
  Z.~G.~Berezhiani and M.~Y.~.Khlopov,
  Sov.\ J.\ Nucl.\ Phys.\  {\bf 52}, 60 (1990)
  [Yad.\ Fiz.\  {\bf 52}, 96 (1990)].
  P.~T.~Winslow, K.~Sigurdson and J.~N.~Ng,
  Phys.\ Rev.\ D {\bf 82}, 023512 (2010)
  [arXiv:1005.3013 [hep-ph]].


\bibitem{Huh:2008vj}
  J.~-H.~Huh, J.~E.~Kim and B.~Kyae,
  Phys.\ Rev.\ D {\bf 79} (2009) 063529
  [arXiv:0809.2601 [hep-ph]];
  M.~Aoki, J.~Kubo and H.~Takano,
  arXiv:1302.3936 [hep-ph].
  A.~Biswas, D.~Majumdar, A.~Sil and P.~Bhattacharjee,
  arXiv:1301.3668 [hep-ph];
  P.~-H.~Gu,
  arXiv:1301.4368 [hep-ph].


\bibitem{Medvedev:2013vsa} 
  M.~VMedvedev,
  arXiv:1305.1307 [astro-ph.CO].
  V.~Semenov, S.~Pilipenko, A.~Doroshkevich, V.~Lukash and E.~Mikheeva,
  arXiv:1306.3210 [astro-ph.CO].

\bibitem{Aprile:2012nq} 
  E.~Aprile {\it et al.}  [XENON100 Collaboration],
  arXiv:1207.5988 [astro-ph.CO].
  

\bibitem{Gondolo:1990dk}
  P.~Gondolo and G.~Gelmini,
  Nucl.\ Phys.\ B {\bf 360} (1991) 145.

\bibitem{Hinshaw:2012aka} 
  G.~Hinshaw {\it et al.}  [WMAP Collaboration],
  arXiv:1212.5226 [astro-ph.CO].

\bibitem{Angloher:2011uu}
  G.~Angloher, M.~Bauer, I.~Bavykina, A.~Bento, C.~Bucci, C.~Ciemniak, G.~Deuter, F.~von Feilitzsch {\it et al.},
  [arXiv:1109.0702 [astro-ph.CO]].
  

\end{thebibliography}
\end{document}